\documentclass{article} 
\usepackage[dvips]{graphicx}
\usepackage{amsfonts,amsmath,amssymb}
\usepackage{hyperref}

\begin{document}

\title{Hawking temperature and the emergent cosmic space}
\author{M. Hashemi$^1$,\, S. Jalalzadeh$^1$\footnote{s-jalalzadeh@sbu.ac.ir}\,\,\,\,and\,\,\,S. Vasheghani Farahani$^2$\\
\\\begin{small}$^1$ Department of Physics, Shahid Beheshti University, G. C., Evin,Tehran, 19839, Iran\end{small}
\\\begin{small}$^2$ Department of Physics, Tafresh University, Tafresh, P.O. Box 39518-79611, Iran\end{small}}

\maketitle
\begin{abstract}
The aim of this work is to model the evolution of the cosmic space based on thermodynamical parameters. The universe is considered to have an apparent horizon radius with a Kodama-Hayward temperature assigned to it. The method is founded on the fact proposed by Padmanabhan \cite{Pad1,Pad2} that the subtraction of the surface and bulk degrees of freedom provides information on the cosmic space emergence. The fact of the matter is that in this approach the Raychaudhuri equation could even be obtained by the consideration of only thermodynamical parameters.  As such, the standard general relativity is taken as the starting point where by implementing the standard cosmological equations we obtain a generalized evolutionary equation supporting emergence of the cosmic space. The method proposed in this work would provide basis for other cosmological models to have an emergent perspective.\\\\
\textbf{Pacs:} 04.20.Cv, 04.50.-h, 04.70.Dy
\end{abstract}
\section{Introduction}
The concept of emergence features itself in all aspects of science. To comply with the aims of the present study we only focus on the emergence linked to gravity, which is named by Sakharov as the induced gravity. This induced gravity is obtained by performing a mean field approximation on the degrees of freedom of spacetime atoms \cite{Sak}.\\
In the context of black holes the concept of surface gravity is referred to the temperature, where this temperature points on the horizon of the black hole. However, it is the area of this horizon that refers to the entropy of the black hole, for details see \cite{BCH,Bek1,Bek2,Haw}. It was not until the studies performed by Jacobson that a robust connection between thermodynamics and gravity was established. The starting point for Jacobson was the Clausius relation ($\delta Q=TdS$) where by implementing the equivalence principle, derived Einstein's field equations proving that they are the equation of state for the spacetime \cite{Jac}. Note that $T$ represents the Unruh temperature \cite{Unr} observed by an observer accelerating just inside the horizon, and $\delta Q$ represents the energy flux across the horizon. \\
The definition of gravity has been elevated from a fundamental interaction to an entropic force. This entropic force is due to the variations of the entropy extracted from the holographic screen \cite{Ver}. This approach enabled Verlinde to express Newton's law of gravity by implementing the entropic force, holographic principle and the equipartition law for energy. In addition, he derived Newton's second law of mechanics using the entropic force and Unruh temperature. It is worth stating other assumptions and methods that enables obtaining the laws of Newton. Padmanabhan obtained Newton's law of gravity by assuming that the equipartition law of energy works for the horizon degrees of freedom by taking $S=E/2T$, where E is an active gravitational mass \cite{Pad3}. In this line, Cai illustrated that the equipartition law of energy for the horizon degrees of freedom, the holographic principle with the Unruh temperature would lead to the Friedmann equations for flat spacetime \cite{Cai4}. By considering the Lovelock gravity with an arbitrary spatial curvature, a modified version for the Friedmann equations has also been obtained \cite{Shey}.\\
An interesting aspect of the Lanczos-Lovelock models came prominent when Padmanabhan stated that the Lagrangian of the bulk and the surface terms have a direct proportionality, illustrated by 
\begin{eqnarray}\label{1}
\sqrt{-g}L_{{sur}}=-\partial_a\left( g_{ij}\frac{\delta(\sqrt{-g}L_{{bulk}})}{\delta(\partial_ag_{ij})}\right).
\end{eqnarray}
It could readily be deduced from Eq. (\ref{1}) that the information possessed by the bulk is similar to that of the surface. Note that the boundary and bulk terms in Eq. (\ref{1}) respectively indicate entropy and energy, see \cite{Pad2}. 
In this context the gravitational action works similar to the free energy of a static spacetime that possesses a horizon.
This statement is born out of the fact that in the context of gravity, the standard action principle is equivalent to the extremum of the free energy of spacetime \cite{Kol}. This statement together with the proposition issued by Verlinde \cite{Ver} enabled Padmanabhan to say \textit{``Cosmic space is emergent as cosmic time progresses"}. Note that Verlinde had treated the concept of time different from space which caused the existence of general covariance to be invalid. Moreover, it is not best to say that space is emergent around finite gravitating systems, e.g. the Sun-Earth system. But Padmanabhan interestingly showed that at least in the context of cosmology, by taking some desired time interval, say cosmic time, the issues raised by Verlinde are out of the question. This is how Padmanabhan came to issue the statement that the expansion of the universe continues till the holographic equipartition takes place. He proposed that in an infinitesimal interval, $dt$, of the cosmic time, the variation of the cosmic volume, $dV$, is given by
\begin{eqnarray}\label{2}
\frac{dV}{dt}=L^2_p(N_{{sur}}-N_{{bulk}}),
\end{eqnarray}
where $N_{sur}$ and $N_{bulk}$ are the surface and bulk degrees of freedom, respectively. The Planck length is represented by $L_P$. Equation (\ref{2}) is the simplest accessible form for describing the asymptotic holographic principle. It is worth naming some interesting proposals based on Padmanabhan's proposal (Eq. (\ref{2})) that have been issued to comply with the physical context under consideration. As such, Cai expanded Eq. (\ref{2}) to (n+1)-dimensions \cite{Cai1} written as
\begin{eqnarray}\label{2Cai}
\alpha \frac{dV}{dt}=L^{n-1}_p(N_{{sur}}-N_{{bulk}}),
\end{eqnarray}
where $\alpha$ is equal to $\frac{n-1}{2(n-2)}$. Sheykhi took in to account the apparent horizon instead of the hubble horizon and wrote \cite{Shey}
\begin{eqnarray}\label{2Shey}
\alpha \frac{dV}{dt}=L^2_p\frac{R_A}{H^{-1}}(N_{{sur}}-N_{{bulk}}),
\end{eqnarray}
where $R_A$ is the apparent horizon radius and $H=\dot a/a$ is the Hubble parameter. {Although this equation works fine, but is however not consistent with Padmanabhan\rq{}s proposal. The reason for this is that the RHS of equation (\ref{2Shey}) is not totally based on thermodynamical quantities. Strictly speaking,  a thermodynamical interpretation of the term ${R_A}/{H^{-1}}$ is not simply achieved.} Eune et al. proposed \cite{Eune}
\begin{eqnarray}\label{2Eune}
  \frac{dV_k}{dt} = L_ P^2 f_k(t) (N_{sur} - N_{ bulk}),
\end{eqnarray}
with
\begin{eqnarray}
f_k(t)=\frac{\tilde{V}_k}{V_k}\left[\frac{\frac{\dot{R}_AH^{-1}}{R_A}-\frac{R_A}{H^{-1}}\frac{V_k}{\tilde{V}_k}+1}
{\frac{\dot{R}_AH^{-1}}{R_A}+\frac{R_A}{H^{-1}}\frac{V_k}{\tilde{V}_k}-1}\right],
\end{eqnarray}
where $V_{k}$ is the volume of the universe, and $\tilde{V}_k$ is equal to $4\pi R_A^3/3$. {Note that $f_k(t)$ is the deviation volume coefficient from flat universe, where a thermodynamic interpretation for it is not simple.} However there are more proposals issued which are out of the scopes of the present work, for details see \cite{Eune,Lingpan,chenshao}. Note that Eqs. (\ref{2Shey} - \ref{2Eune}) all contribute towards providing an insight on the thermodynamical effects. But, an expression based only on thermodynamical parameters still lacks. The reason that we make this statement is that the parameters, $\dot {R}_A$ and $H$ that feature in Eqs. (\ref{2Shey} - \ref{2Eune}) are not thermodynamical parameters. Therefore, we intend to present a dynamical emergent equation based only on thermodynamical  parameters. {Now by implying an apparent horizon (which is considered as the most appropriate boundary in application to thermodynamics) alongside Kodama temperature (which is a temperature for an evolving horizon implied in cosmology) will enable us to write the RHS of the dynamical emergent equation based only on thermodynamical parameters.}
\par In the next section we discuss the thermodynamics of the apparent horizon. In section three, we propose a generalised version for the Raychaudhuri equation based on thermodynamical parameters. In addition to the Einstein gravity, the generalisation is performed in (n+1)-dimensions. In the final section we provide a list of the emergent proposals based on the holographic Raychaudhuri equation which comes from the Einstein general relativity. We also investigate the necessary considerations for one to build a proposal leading to the Raychaudhuri equation.
\section{Thermodynamics of the Apparent Horizon}
Consider a spatially homogeneous and isotropic Friedmann-Lema\'{\i}tre-Robertson-Walker (FLRW) universe. Its line element is expressed by
\begin{eqnarray}\label{3}
ds^2=h_{\mu\nu}dx^{\mu}dx^{\nu}+R^2(d\theta^2+\sin^2\theta d\phi^2),\,\,\,\,\,\,\,\,\,\,(\mu,\nu=0,1 ),
\end{eqnarray}
with $R=a(t)r$, $x^0=t$, $x^1=r$. Note that the two dimensional metric $h_{\mu\nu}$ is equal to ${diag}(-1, a^2/(1-kr^2))$, where $k$ is $\pm1$ or 0, corresponding to the closed, open or flat universes.  In such a condition and by taking $h^{\mu\nu} \partial_\mu R\partial_\nu R$ equal to zero, the dynamical apparent horizon which is the only horizon that enables the consistency of all thermodynamical properties over it,  is expressed by \cite{Cai2,Far}
\begin{equation}\label{4}
R_A=\left(H^2+\frac{k}{a^2}\right)^{-\frac{1}{2}}.
\end{equation}
Note that the dynamical apparent horizon is the boundary of a trapped surface with no expansion $\theta_l \theta_n=0$ , where the indexes $l$ and $n$  respectively indicate the outgoing and ingoing radial null geodesics  \cite{Far}.\\
It should be stated that the Hubble horizon  ($R_H=\frac{1}{H}$)  only gives the order of magnitude for the radius curvature of the FLRW. That is fine for the estimation of the event horizon radius during inflation when the universe is close to a de Sitter space \cite{Far}. But when talking about local issues, the Hubble horizon due to its global considerations might not be the best choice. Hence to comply with the aims of the present study we implement the apparent horizon. The locality of the apparent horizon is due to its definition based on the  implementation of null geodesic congruences and their expansions, with no reference to the global causal structure. The apparent horizon provides a surface where due to the black hole dynamics brings thermodynamics in to play.  As such, a surface gravity be associated to the apparent horizon  
\begin{eqnarray}\label{5}
\kappa=\frac{1}{2\sqrt{-h}}\partial_\mu\left(\sqrt{-h}h^{\mu\nu}\partial_\nu R\right)
=-\frac{1}{R_A}\left(1-\frac{\dot{R}_A}{2HR_A}\right).
\end{eqnarray}
For details see (\cite{Far} - \cite{Wei}). The working temperature in the present study is the Hawking temperature (Kodama-Hayward temperature) which is obtained for the apparent horizon as
\begin{eqnarray}\label{9}
T_A=\frac{|\kappa|}{2\pi}
=\frac{\pm1}{2\pi R_A}\left(1-\frac{{\dot {R}}_A}{2HR_A}\right).
\end{eqnarray}
The reason for implementing this temperature is due to the fact that here the selected boundary for our universe is the apparent horizon. We will show that this choice for the temperature works fine in the context of emergent cosmology \cite{Cai2, Far, SZ}.
Note that the choice for the sign of the unity on the RHS of Eq. (\ref{9}) is to  prevent the expression on the RHS to be negative. In case of selecting the plus sign, the heat capacity of the universe would be positive, which implies the universe to be thermodynamically stable \cite{Akbar}.

\par In studies prior to this work by considering the time interval very small, the radius of the apparent horizon was taken fixed \cite{Shey,Eune,Cai2,Cai5}. This enabled them to neglect the second term on the RHS of Eq. (\ref{9}).  However, this may not comply to the fact that the cosmic space is emerging with time. In addition, since the aim is to obtain a dynamical emergent equation for $R$, neglecting $\dot R$ in Eq. (\ref{9}) seems irrelevant. To comply with the properties of the emerging cosmic space, we would not neglect the second term in the parenthesis on the right hand side of Eq. (\ref{5}). By neglecting it we would have obtained $\kappa\approx -(R_A)^{-1} $, which is out of the question for the present work. Now going down to facts, for the pure de Sitter universe, $N_{{sur}}$ is equal to $N_{{bulk}}$ (\cite{Pad1,Pad2}), where our universe due to the late time acceleration can be considered asymptotically de Sitter. As such, the surface and bulk degrees of freedom may no longer equal each other. In fact, the key idea of the present work is that the difference between the degrees of freedom on the surface and inside the bulk changes the apparent volume of the universe, disabling the the consideration of a constant radius for the horizon. This results in what we call the apparent volume sphere (volume of cosmic sphere), to evolve. The process of evolution is directed so how to maintain the equality of the surface and bulk in the sense of degrees of freedom.  To state clearer, the tendency is to finally find a holographic equipartition for the universe.  To this end we need to rewrite the Raychaudhuri equation based on the thermodynamical parameters named earlier in this section.

\section{Friedmann and Raychaudhuri equations}
The standard model of cosmology is based  on the perfect fluid description of the universe which is constructed by three equations. Two equations can be taken as reference in order to produce the third. However, this choice depends on the  intentions of the study. In the present study the two reference independent equations are
\begin{eqnarray}\label{6}
H^2+\frac{k}{a^2}=\frac{1}{R_A^2} =\frac{8\pi L_p^2}{3}\rho,
\end{eqnarray}
which is the Friedmann equation\footnote{For simplicity, the natural units $k_{{B}}=c=\hbar=1$ are used throughout this paper.}, and 
\begin{eqnarray}\label{7}
\dot{H}+H^2=-\frac{4\pi L_p^2}{3}(\rho+3p),
\end{eqnarray}
which is the Raychaudhuri equation. The parameters  $\rho$ and $p$ respectively represent the total energy density and total matter pressure of the universe. Note that the dot over $H$ denotes its time derivative. The equation which is born out of the Friedmann and Raychaudhuri equation is  the continuity equation as of the form 
\begin{eqnarray}\label{8}
\dot{\rho}=-3H(\rho+p).
\end{eqnarray}
By differentiating Eq. (\ref{6}) in respect to cosmic time, we obtain
\begin{eqnarray}\label{10}
\frac{4\pi L_p^2}{3}\dot{\rho}=-\frac{\dot{R_A}}{R_A^3},
\end{eqnarray}
where by combining Eqs. (\ref{9}) ,(\ref{6}) ,(\ref{8}) and (\ref{10}) we obtain
\begin{eqnarray}
\frac{{\dot {R}}_A}{2HR_A}&=&1-2\pi R_AT_A\label{11}=\frac{3}{4}(1+\frac{p}{\rho}).\label{12}
\end{eqnarray}
Equation (\ref{12}) clearly shows that the term $\dot{R}_A/2HR_A$ except for the case of dark energy ($p_{de}=-\rho_{de}$) which is equal to zero, is not negligible on the RHS of Eq. (\ref{9}). This is due to the fact that for the dust or radiation universe the term $\dot{R}_A/2HR_A$ is equal to $3/4$ and $1$ respectively, making it non-negligible compared to unity.  However the term $\dot{R}_A/2HR_A$ is more pronounced in regions far from the applicability of the holographic principle. By combining Eqs. (\ref{9}) and (\ref{12}) the exact expression for the temperature is obtained
\begin{eqnarray}\label{13}
T_A=\frac{1}{2\pi R_A}\left(1-\frac{{\dot {R}}_A}{2HR_A}\right)=\frac{1}{8\pi R_A}\left(1-3\frac{p_{\rm{tot}}}{\rho_{\rm{tot}}}\right).
\end{eqnarray}
Equation (\ref{13}) is the general formula for the horizon temperature. It is worth noting that in emergent cosmology due to the asymptotic de sitter behavior the existence of dark energy is compulsory, see \cite{Pad2}. Thus, the temperature will never be zero. {This means that the total energy density, $\rho$,  would never be equal to triple of the total pressure, 3p. Hence the temperature would never become zero.} Now that the general relation for the horizon temperature (Eq. (\ref{13})) is obtained,  we have found ourselves a starting point to reproduce the Raychaudhuri equation in terms of the surface degrees of freedom ($N_{sur}$), the bulk degrees of freedom ($ N_{bulk})$, volume, and temperature. But first we need to state some definitions. The number of surface degrees of freedom has been proposed by Padmanabhan as \cite{Pad1}
\begin{eqnarray}\label{14}
N_{sur}=\frac{A}{L^{2}_{P}}=\frac{4\pi R_A^{2}}{L^{2}_{P}},
\end{eqnarray}
where $A=4\pi R_A^{2}$ denotes the cosmic sphere surface. The Komar energy contained inside the bulk is \cite{Pad4}
\begin{eqnarray}\label{15}
E_{{Komar}}=|(\rho+3p)V_A|=-\varepsilon(\rho+3p)V_A,\quad\quad\quad\quad
\varepsilon \equiv\begin{cases}
+1 & \rm{for}\quad \frac{P}{\rho}<\frac{-1}{3} ,\\
-1 & \rm{for}\quad \frac{P}{\rho}>\frac{-1}{3}, \\
\end{cases}
\end{eqnarray}
where $V_A=\frac{4}{3}\pi R_A^3$. Note that for an accelerated universe we have $\varepsilon=-1$, where for a decelerated  universe we have $\varepsilon=+1$ . The bulk degrees of freedom which obeys the equipartition law of energy has been defined in  \cite{Pad2} as
\begin{eqnarray}\label{16}
N_{{bulk}}=\frac{2}{T_A}E_{{Komar}}.
\end{eqnarray}
Now by differentiating Eq. (\ref{6}) with respect to  time we obtain
\begin{eqnarray}\label{17}
\dot{H}-\frac{k}{a^2}=-\frac{\dot R_A}{HR_A^3},
\end{eqnarray}
where by implementing Eq.  (\ref{11}) one can write 
\begin{eqnarray}\label{18}
\dot{H}-\frac{k}{a^2}=-\frac{2}{R_A^2}(1-2\pi R_A T_A).
\end{eqnarray}
Owing to the definition of the apparent horizon (Eq. (\ref{4})), Eq. (\ref{18}) can be simplified to
\begin{eqnarray}\label{19}
\dot{H}+H^2=-\frac{1}{R_A^2}+\frac{4\pi T_A}{R_A}.
\end{eqnarray}
By combining Eqs. $(\ref{7})$ and ($\ref{19}$), we obtain
\begin{eqnarray}\label{20}
-2R_A+4\pi R_A^2T_A=-4\pi R_A^2T_A-\frac{8\pi R_A^3 L_p^2}{3}(\rho+3P).
\end{eqnarray}
Substituting Eq. (\ref{12}) with its equivalent on the LHS of Eq. (\ref{20}) we have 
\begin{eqnarray}\label{21}
\begin{array}{cc}
-\frac{\dot V_A}{4\pi R_A^2H}=L_p^2 T_A (-N_{sur}+\varepsilon N_{bulk}),
\end{array}
\end{eqnarray}
 In order to write Eq. (\ref{21}) in a more conclusive form we write
\begin{eqnarray}\label{23}
\frac{dV}{dt}=\frac{2}{T^2_p}\frac{T_A}{T_H}(N_{sur}-\varepsilon N_{bulk}),
\end{eqnarray}
{
where $T_p$ is the Planck temperature. The right hand side of the dynamical equation (\ref{23}) indicates that the universe tends to the holographic principle, asymptotically. As stated earlier, in order to measure the temperature of the evolving horizons it is suitable to use the Kodoma temperature, $T_A$ , which is measured by the Kodama observer. While, using the measured temperature by a comoving observer, $T_H=(2\pi HR_A^2)^{-1}$ \cite{Hu}, for the cosmological horizon \cite{Hu} proves conclusive. 
It is known that various observers do not measure equal temperatures for evolving horizons. However it is only for a flat de Sitter universe that the two observers record similar readings. This is when the the cosmological horizon equals the apparent horizon\cite{Hu}. Hence, in the emergent cosmology, when the universe tends to the de Sitter universe, the ratio of $T_A/T_H$ tends to unity resulting in similar measurements provided by the two observers. }
The interesting aspect of Eq. (\ref{23}) is that it provides a general form for the dynamical emergent description of the cosmic space based on information and thermodynamical parameters in (3+1)-dimensions.
\par In this stage we intend to extend the applicability of the dynamical emergent equation to an (n+1)-dimensional spacetime. In this line, we recall the Friedmann and Raychaudhuri equations together with the continuity equation for the $(n+1)$-dimensional homogeneous and isotropic universe which are 
\begin{eqnarray}
\begin{array}{cc}
H^{2}+\frac{k}{a^2}=\frac{1}{R_A^2}=\frac{16\pi L^{n-1}_P}{n(n-1)}\rho,\label{24}
\\\\\dot H+H^2=\frac{-8\pi L^{n-1}_{P}}{n(n-1)}[(n-2)\rho+np)],\label{25}
\end{array}
\end{eqnarray}
and
\begin{eqnarray}\label{26}
\dot \rho=-nH(\rho+p).
\end{eqnarray}
Following the same procedure as in the case for (3+1)-dimensions, we obtain the time dynamical emergent equation in (n+1)-dimensions 
\begin{eqnarray}\label{38}
\frac{dV_n}{dt}=\frac{2}{\alpha T_p^{n-1}}\frac{T_A}{T_H}(N_{sur}-\varepsilon N_{bulk}),
\end{eqnarray}
where $T_p$ is the Planck temperature in (n+1)- dimensions. Equation (\ref{38}) is the holographic Raychaudhuri equation for arbitrary dimensions. Note that the apparent horizon temperature and surface degrees of freedom in (n+1)-dimensions has been considered as 
\begin{eqnarray}\label{31}
T_A=\frac{1}{2\pi R_A}\left(1-\frac{{\dot {R}}_A}{2HR_A}\right)
=\frac{1}{8\pi R_A}\left(4-n-n\frac{p}{\rho}\right),
\end{eqnarray}
and 
\begin{eqnarray}\label{32}
N_{{sur}}=\frac{\alpha A}{L^{n-1}_{P}},
\end{eqnarray}
respectively. Where $A=n \Omega_{n}R_A^{n-1}$, $\alpha=\frac{n-1}{2(n-2)}$ and $\Omega_{n}$ is the volume of the unit $n$-dimensional sphere \cite{Ver}. In addition the $(n+1)$-dimensional form of the Komar energy has been taken as \cite{Ver,Cai4,BR}
\begin{eqnarray}\label{33}
E_{{Komar}}=\left|\frac{(n-2)\rho+np}{n-2}V_n\right|=-\varepsilon \frac{(n-2)\rho+np}{n-2}V_n, \quad\quad\quad 
\varepsilon \equiv\begin{cases}
+1 & \rm{for}\quad \frac{P}{\rho}<\frac{2-n}{n}, \\
-1 & \rm{for}\quad \frac{P}{\rho}>\frac{2-n}{n}, \\
\end{cases}
\end{eqnarray}
where the volume $V$ is equal to $\Omega_{n}R_A^n$. Note that the bulk degrees of freedom for (n+1)-dimensions is the same as for the (3+1)-dimensions, given by Eq. (\ref{16}). \\
It is instructive to list the proposals having a mutual interest with the present work which have been dedicated to the dynamic emergent equation based on Holographic Raychaudhuri's model in (3+1)-dimensions, see table (1). In table (1) some of the most important dynamical emergent equation proposals based on the holographic Raychaudhuri's equation (\ref{38}) are revisited which enables a better understanding on their facts and features. 
\begin{table}[htb] \label{table1}
\caption{The listed proposals with their applicability for the dynamical emergent equation}
\centerline{\begin{tabular}{|l||ccccc|}
 \hline
  & Padmanabhan & Cai & Sheykhi& Eune et al.  &  Holographic \\
 & \cite{Pad1} & \cite{Cai1} &\cite{Shey}& \cite{Eune}  & Raychaudhuri \\
 \hline
 \hline
  &  &  &  & \\
\(\left[L^2_p(N_{sur}-N_{bulk})\right]^{-1}\frac{dV}{dt}\)& 1& 1 &\(\frac{R_{A}}{H^{-1}}\) &\(f_k(t)\) &\(2\frac{T_A}{T_H}\) \\
 &  &  &  & \\
 \hline
cosmic boundary& hubble & hubble & apparent &apparent &apparent \\
 & horizon & horizon & horizon & horizon & horizon\\
\hline
dimension  & 3+1 & n+1 & n+1 & n+1& n+1\\
\hline
Hawking temperature  & approximate due & approximate due& approximate due & approximate due & exact due to\\
 & to neglection of  & to neglection of & to neglection of & to neglection of & keeping the\\
            & the second term & the second term & the second term  & the second term& second term\\
\hline
horizon radius &constant  &constant  &constant  &constant &dynamic \\
\hline
only depends on the &  &  &  & \\
thermodynamical  & Yes & Yes & No& No & Yes  \\
quantities &  &  &  &  \\
 \hline
\end{tabular}}
\end{table}
\section{Concluding Remarks}
In this work we have studied cosmology from the view point of thermodynamics. In this line the Raychaudhuri equation has been reproduced by considering the emergence of cosmic space. This has been provided by taking in to account the following considerations
\begin{itemize}
\item Dynamical equations appear only in case of a dynamic horizon. It is well known that the emergence of a cosmic space is in contradiction with a constant horizon radius.
\item The hard core of such theories is due to thermodynamics and information. Hence, there is growing demand for the explanation of the dynamical emergent equation in terms of thermodynamical quantities.
\item From the view point of thermodynamics, the most reasonable boundary for the universe is the apparent horizon. This is due to the fact that the apparent horizon is the only horizon that all thermodynamic laws apply on it. 
\end{itemize}
It has been proposed by Padmanabhan that the emergence of the cosmic space and expansion of the universe are due to the difference between the number of degrees of freedom on the horizon surface and in the emerged bulk (Eq. (\ref{2})). This statement is true, but one must mind that the emergence of the cosmic space does not permit neglecting the second term on the RHS of Eq. (\ref{9}). This is due to the fact that the second term on the RHS of Eq. (\ref{10}) is time dependent, therefore, it is best not to neglect it. Regarding this issue the Raychaudhuri equation has been rewritten to illustrate the functionality between $N_{sur}$ and $N_{bulk}$ in terms of thermodynamics and information. Note that in the emergent scenario, first of all, one must select a line element for the universe in order to obtain $R_A$ and $T_A$ by implementing the trapped surface equation and Hawking temperature, respectively.\\ 
This brings need for a proposal that only depends on the thermodynamical quantities besides a dynamical horizon. Therefore in order to describe our universe as an emerging cosmic space by the Raychaudhuri equation, we have obtained Eq. (\ref{38}) which is for (n+1)-dimensions.\\
Finally, we must state that due to the ability of the Raychaudhuri and continuity equations for describing our universe, we provided a robust form of the Raychaudhuri equation in terms of thermodynamical or informational parameters. This work provides basis for studying the continuity equation and rewrite it in terms of informational parameters. This will be performed in a follow up paper.



\end{document}